# Exciton spin relaxation in InAs/InGaAlAs/InP(001) quantum dashes emitting near 1.55 μm

M. Syperek,[1,a)] Ł. Dusanowski,[1] M. Gawełczyk,[1,2] G. Sęk,[1] A. Somers,[3] J. P. Reithmaier,[3,4] S. Höfling,[3,5] and J. Misiewicz[1]

[1]*Laboratory for Optical Spectroscopy of Nanostructures, Department of Experimental Physics, Faculty of Fundamental Problems of Technology, Wrocław University of Science and Technology, Wybrzeże Wyspiańskiego 27, 50-370 Wrocław, Poland*
[2]*Department of Theoretical Physics, Faculty of Fundamental Problems of Technology, Wrocław University of Science and Technology, Wybrzeże Wyspiańskiego 27, 50-370 Wrocław, Poland*
[3]*Technische Physik, University of Würzburg and Wilhelm-Conrad-Röntgen-Research Center for Complex Material Systems, Am Hubland, D-97074 Würzburg, Germany*
[4]*Institute of Nanostructure Technologies and Analytics (INA), CINSaT, University of Kassel, Heinrich-Plett-Str. 40, 34132 Kassel, Germany*
[5]*SUPA, School of Physics and Astronomy, University of St. Andrews, North Haugh, KY16 9SS St. Andrews, United Kingdom*

Exciton spin and related optical polarization in self-assembled InAs/In$_{0.53}$Ga$_{0.23}$Al$_{0.24}$As/InP(001) quantum dashes emitting at 1.55 μm are investigated by means of polarization- and time-resolved photoluminescence, as well as photoluminescence excitation spectroscopy, at cryogenic temperature. We investigate the influence of highly non-resonant and quasi-resonant optical spin pumping conditions on spin polarization and spin memory of the quantum dash ground state. We show that a spin pumping scheme, utilizing the longitudinal-optical-phonon-mediated coherent scattering process, can lead to the polarization degree above 50%. We discuss the role of intrinsic asymmetries in the quantum dash that influence values of the degree of polarization and its time evolution. *Published by AIP Publishing.* [http://dx.doi.org/10.1063/1.4966997]

Self-assembled InAs quantum dashes (QDashes), epitaxially grown on an InP(001) substrate, resemble quantum dots (QDs), however, strongly elongated in one of the in-plane dimensions.[1–4] So far, such QDashes have been exploited mostly as a gain medium in semiconductor lasers, amplifiers, or superluminescent diodes suited for telecom technology operating at 1.3 and 1.55 μm low-loss spectral windows of silica fibers.[5–7] Recent research promises possible applications of QDashes in long-haul secure quantum data transmission lines.[8,9] An InAs/InP(001) QDash-based non-classical single photon emitter operating at 1.55 μm has been demonstrated[8] along with a possibility to tune the exciton fine structure splitting down to zero by applying an external magnetic field.[9] While the former demonstrates a capability of QDashes to generate a single photon at a time, the latter can lead to generation of polarization-entangled photon pairs at telecom wavelengths, essential for, e.g., a quantum repeater technology. Since semiconductor QDashes can be considered as a bridge platform between a solid-state quantum information storage/operation and a quantum state of light, it is crucial to investigate properties of confined spin states that can mediate an exchange of quantum information.

The effects concerning spin excitation and spin-related phenomena in self-assembled quasi-0D quantum systems capable of generating photons at 1.55 μm wavelength have not been investigated very extensively so far. Existing reports address only the problem of either exciton or electron/hole g-factors in InAs/InP QDs.[10–13] However, issues such as the longitudinal or transverse spin relaxation or the role of spin pumping schemes on the spin memory effect in this particular quantum system have not been explored up to date.

In this letter, we investigate properties of polarized emission and spin states of excitons confined in an ensemble of InAs/In$_{0.53}$Ga$_{0.23}$Al$_{0.24}$As/InP(001) QDashes by means of polarization- and time-resolved photoluminescence (TRPL), as well as photoluminescence excitation spectroscopy (PLE). We demonstrate various schemes of spin injection and their impact on the spin memory effect in QDashes emitting near 1.55 μm.

The investigated sample was grown in an EIKO gas source molecular-beam epitaxy system on a sulfur-doped InP(001) substrate. The structure consists of QDashes formed in the Stranski-Krastanov growth process by a deposition of InAs layer of nominal thickness of 1.3 nm at 470 °C onto a 200 nm-thick In$_{0.53}$Ga$_{0.23}$Al$_{0.24}$As barrier. QDashes were covered by 100 nm of In$_{0.53}$Ga$_{0.23}$Al$_{0.24}$As and the layer sequence was finalized by a 10 nm-thick InP cap layer. Both barriers are lattice-matched to InP and were grown at 500 °C. Structural data reveal that QDashes are triangular in a cross-section, with 20 nm in base width and 3.5 nm in height. The length varies between 50 and hundreds of nanometers. The areal density of QDashes is ~$5 \times 10^{10}$ cm$^{-2}$. QDashes are nominally undoped, however, a small residual electron doping may be present.

For time-integrated photoluminescence (PL) and TRPL experiments, the structure was held in a continuous flow liquid helium cryostat at $T = 4.2$ K and was illuminated through a microscope objective (NA = 0.4) by a train of laser pulses, with a pulse duration of ~2 ps and 76 MHz repetition

a)Electronic mail: marcin.syperek@pwr.edu.pl

frequency. In the case of resonant excitation conditions, the pulse train was generated by an optical parametric oscillator, synchronously pumped by a mode-locked Ti:Sapphire laser. This system provides a tunability of the photon energy in the range of 0.82–1.24 eV (1–1.51 $\mu$m). In highly non-resonant excitation conditions, only the Ti:Sapphire laser operating at 1.49 eV (0.83 $\mu$m) photon energy was used. Photons emitted from the structure were collected by a microscope objective and directed to a spectral analyzer consisting of a 0.3 m-focal length monochromator and an InGaAs-based multichannel detector or a state-of-the-art nitrogen-cooled streak camera system from Hamamatsu, operating in a photon counting mode. The streak camera system covers a spectral range of 1.0–1.7 $\mu$m and provides a temporal resolution on the level of ~20 ps. The linear polarization of excitation was controlled by a calcite polarizer with the extinction ratio of $10^5$:1 and a multi-order half-wave plate. The polarization of emission was analyzed in front of the monochromator, where the incident light passes through a multi-order half-wave plate first, and then through a calcite polarizer set in a fixed position. This allowed us to eliminate a possible impact of the internal elements of the monochromator on the light analysis process.

A low temperature PL spectrum obtained from the ensemble of studied InAs/In$_{0.53}$Ga$_{0.23}$Al$_{0.24}$As/InP(001) QDashes is shown in Fig. 1(a). It was collected under the non-resonant excitation, in which e-h pairs are photogenerated mainly in the In$_{0.53}$Ga$_{0.23}$Al$_{0.24}$As barrier and in the InP capping layer and subsequently populate the QDash states after the energy dissipation and total angular momentum relaxation. Since the excitation was rather weak (~1 e-h pair/QDash), the observed PL spectrum is produced mainly by the recombination of confined neutral excitons, possibly partially affected by the presence of negatively charged excitons (trions) and biexcitons. The observed ~35 meV broadening of the PL band reflects the ensemble non-uniformity caused by fluctuations in, e.g., QDash size, strain and chemical composition. Most of the QDashes are preferentially aligned and elongated along the [1$\bar{1}$0] crystallographic direction, with the in-plane aspect ratio exceeding 2.5. Despite other effects, such a geometrical property of a QDash, especially the lack of in-plane rotational symmetry, can already suggest existence of polarization anisotropy in the light emission from the QDash ground state (GS),[14] and hence it must be addressed before the analysis of any spin properties of the system.

In order to examine the optical anisotropy, the structure was excited in the barrier by linearly polarized pulses with two directions of polarization axis: V ([1$\bar{1}$0])—along the QDash, and H ([110])—in a perpendicular direction (see the inset in Fig. 1(a)). Two linearly polarized components of emission, labeled as $I_V$ and $I_H$, were measured with respect to these directions. The results collected in Fig. 1(b) confirm strong polarization anisotropy of the emission process as the two cross-polarized PL spectra exhibit a significant difference in their peak intensities. For a quantitative discussion, one can introduce the degree of linear polarization, $DOP_{V(H)} = \frac{I_{V(H)} - I_{H(V)}}{I_{V(H)} + I_{H(V)}}$. In the above-mentioned case, the $|DOP_{V(H)}|$ reaches $25 \pm 5\%$ which defines the so-called intrinsic $DOP$. It is important to note that the intrinsic $DOP$ is not a result of the process of building-up a certain exciton spin population of the GS in the QDash ensemble. The non-resonant excitation results in energy dissipation accompanied by efficient spin relaxation that in turn erases the memory of an initial exciton spin state acquired from the polarization of excitation. In these conditions, a substantial $DOP$ appears due to intrinsic properties of the e-h confinement that pins the polarization state of emission. This can be explained in terms of the QDash shape, anisotropic confinement for carriers, a non-uniform strain field and piezoelectricity induced by it, atomistic disorder at interfaces, and finally the local asymmetry of the InAs zinc-blende crystal lattice.[10,14–18]

From a theoretical point of view, these asymmetries with respect to V and H axes lead to a light-hole (lh, $|\Downarrow/\Uparrow\rangle$) admixture to a nominally purely heavy-hole ($|\Uparrow/\Downarrow\rangle$) state, producing hole eigenstates,[19] i.e., $|\Uparrow'/\Downarrow'\rangle \propto |\Uparrow/\Downarrow\rangle \pm i\varepsilon|\Downarrow/\Uparrow\rangle$. This converts excitons' polarizations from circular to elliptical, with major axes tilted towards the V axis in the case of both states. This approximately holds for trions due to the lack of an e-h exchange interaction and prevents them from being efficiently addressed with linearly polarized light. In the case of neutral excitons, the presence of an e-h exchange interaction lifts the degeneracy between states. For an in-plane anisotropy introduced between V and H axes, it produces two bright, $|H/V\rangle \propto |\Downarrow\Uparrow'\rangle \pm i|\Uparrow\Downarrow'\rangle$, and two dark eigenstates, $|D1/2\rangle \propto |\Uparrow\Uparrow'\rangle \pm |\Downarrow\Downarrow'\rangle$. Calculations of interband dipole moments,[20] $d_{inter}$, for the bright states indicate that

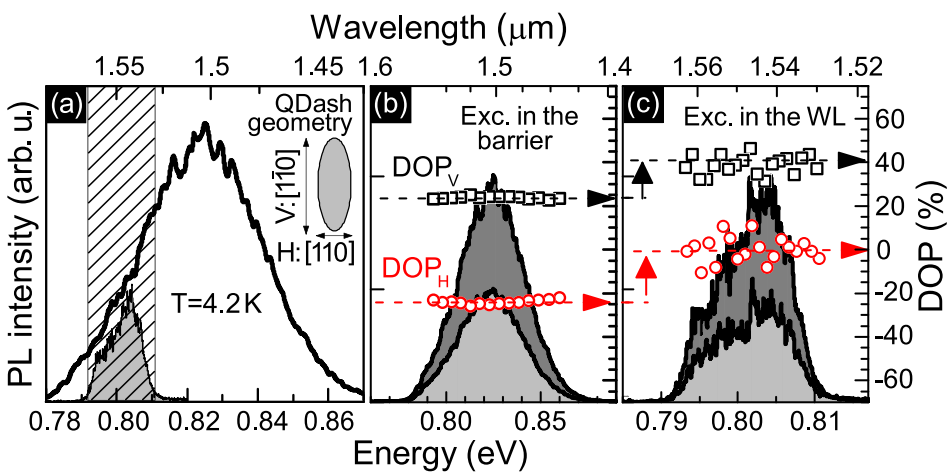

FIG. 1. (a) A low temperature ($T = 4.2$ K) PL spectrum of InAs/In$_{0.53}$Ga$_{0.23}$Al$_{0.24}$As/InP(001) QDashes ($E_{exc} = 1.49$ eV, $P_{exc} = 0.4$ $\mu$W/$\mu$m$^{-2}$). Shaded area shows the spectral range of the bandpass filter. (b), (c) The degree of linear polarization ($DOP$) for the two in-plane directions (H, V) obtained under the pulsed excitation in the barrier ($E_{exc} = 1.49$ eV), and in the WL ($E_{exc} = 1.1$ eV), respectively.

they should couple to light linearly polarized along H and V axes, respectively, with unequal oscillator strengths ($f_V/f_H \propto \varepsilon$ for small lh admixtures). It is considered to be the reason for a substantial value of the intrinsic DOP of the QDash GS emission.

Since the intrinsic DOP is known, one can measure an excess DOP that may result from a certain optical spin pumping scheme. In the following case, the excitation is tuned in energy to the wetting layer (WL) of the structure to decrease the role of spin scattering events leading to a collapse of the desired spin state. Excitons are expected to be created in one of the earlier defined states ($|H\rangle$ or $|V\rangle$) by setting up certain polarization of excitation. Trion states are also partially addressed, however, as their emission is polarized independently of the excitation, it should not contribute to the excess DOP, but act as a background which weakens the observed effect. The emission was analyzed in a similar way as in the highly non-resonant excitation case, however, the PL spectrum was filtered out around 0.80 eV, which allowed for elimination of the scattered laser light. The $DOP_{V(H)}$ is plotted with open points in Fig. 1(c). One may notice that under such an optical pumping scheme, the DOP increased up to $\sim$40% and $\sim$0% in the case of $DOP_V$ and $DOP_H$, respectively. This leads us to a conclusion that the injected spin state is partially preserved during carriers' relaxation to the QDash GS, so the memory of the excitation is present to some extent.

Further, we verified an even more reliable spin pumping procedure, in which the excitation is quasi-resonant with the QDash GS, namely, the pumping via a longitudinal optical (LO) phonon was used to assure the minimal excess energy and nearly immediate injection of a spin state to the QDash GS. Fig. 2(a) presents a 2D map of the $DOP_V$ obtained by scanning the laser photon energy ($E_{exc}$) towards the QDash emission band defined by the spectral filter. One can notice a well resolved intensity feature across the detection energy ($E_{det}$) that shifts parallel to the $E_{exc}$ while keeping a constant distance of $\sim$30 meV (close to the LO phonon energy in InAs) between the excitation and its characteristic energy. For this experiment, the sample was patterned by a $5 \times 5\,\mu$m mesa structure. The discrete character of the map along its vertical axis corresponds to the fact that with a small ensemble size within the mesa and under the quasi-resonant excitation, single emission lines from individual QDashes can be seen. Figures 2(b) and 2(c) present examples of horizontal profiles, registered for the case of V and H pumping, respectively, which were cut-out from the 2D map at two different $E_{det}$. As may be easily noticed, at the 1LO-phonon feature, the DOP is strongly enhanced by $\sim$20% with respect to the DOP measured for the WL excitation and $\sim$35% as compared to the intrinsic DOP. It leads us to an initial conclusion that the creation of exciton, accompanied by emission of a LO phonon, can significantly preserve coherence within the injected spin state as it is likely realized by a coherent inelastic Stokes Raman scattering process. Besides the LO phonon one, a transverse-optical-phonon excitation is possibly observed for about 5–7 meV lower values of detuning (visible also in the PLE map). We skip it in the discussion since it is partly unclear as observed maxima do not correspond to a well fixed value of detuning.

Although the spin memory effect is clearly present, the DOP value is expected to be much higher, up to the limit of 100%.[21] The lack of full polarization of emission in the V-V configuration could be partially caused by (i) random deviations from the V axis orientation in the ensemble of QDashes, (ii) presence of more symmetric structures in the ensemble, (iii) local widenings along QDashes acting as more symmetric trapping centers,[18] (iv) elliptically polarized emission from trions[8] independent of polarization of excitation, acting as a background for excitonic emission.

A more unexpected issue which needs to be addressed here is the asymmetry between V-V and H-H configurations, manifested in a significantly lower values of $DOP_H$. This phenomenon has not been fully understood yet, however, we propose an initial explanation. Based on the preliminary discussion of excitonic states, polarization injection might be expected to be equally effective for V and H cases as interband dipole moments of the two bright states are collinear with respective axes. The interband contribution to the dipole moment is commonly regarded as dominant,[22] however, the usually neglected intraband term may become substantial for structures of a large volume. The macroscopic character of this contribution ($\boldsymbol{d}_{intra} \propto \langle\psi_e|\boldsymbol{r}|\psi_h\rangle$, where $|\psi_{e/h}\rangle$ are the e/h envelope functions), in combination with a significant elongation of QDashes, promotes the V component of $\boldsymbol{d}_{intra}$ approximately to the same extent for both bright states. Such contribution strongly affects the polarization properties of the $|H\rangle$ state as it is perpendicular to its $\boldsymbol{d}_{inter}$, which is not the case for the $|V\rangle$ state. For $|\boldsymbol{d}_{intra}|/|\boldsymbol{d}_{inter}| \sim 1/2$ we were able to approximately reproduce DOP values for both cases of excitation: quasi-resonant ($DOP_V \approx 48\%$, $DOP_H \approx 12\%$) and non-resonant ($DOP_{V/H} \approx \pm 21\%$).

Let us shift the discussion towards dynamic properties of the spin excitation in QDashes. In Figure 3(a), we present TRPL traces registered at $E_{det} = 799.4$ meV under the exciton-1LO phonon spin pumping scheme. As predicted from the lh admixture considerations, the obtained PL lifetimes ($\tau \propto f^{-1}$) for both states slightly differ: $\tau^{V-V} = 1.1 \pm 0.1$ ns vs. $\tau^{H-H} = 1.4 \pm 0.1$ ns. The average exciton lifetime is surprisingly low as compared to $\sim$2.0 ns (Ref. 23) expected for the strong confinement limit. In this case, one has to take into account a dense structure of hole states induced by the QDash size, as indicated by calculations of the QDash band structure,[18,19,23] that influences the exciton radiative lifetime.

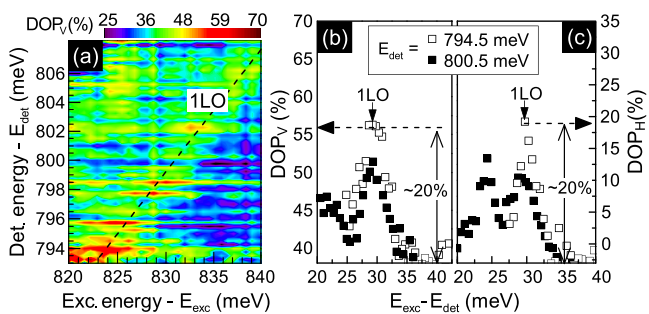

FIG. 2. (a) A 2D map of the $DOP_V$ measured in the exciton-LO phonon excitation scheme at $T = 4.2$ K. Dashed line indicates position of the LO resonance across the QDash emission. (b), (c) vertical profiles of the 2D map at given detection energies indicating enhancement of the DOP at the LO phonon resonance for the $|V\rangle$ and $|H\rangle$ spin state pumping, respectively.

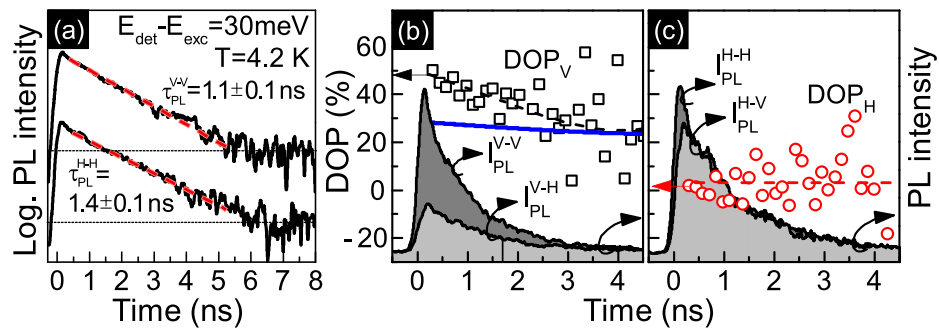

FIG. 3. (a) Low temperature TRPL traces measured within the exciton-1LO phonon excitation scheme for two configurations of the linear polarization of excitation and detection: V-V (upper) and H-H (lower). Dashed red lines are exponential fits. (b), (c) TRPL traces for various excitation-detection schemes: V-V, V-H, H-H, and H-V, and a resultant temporal evolution of $DOP_V$ (black squares in (b)) and $DOP_H$ (red circles in (c)). The blue solid line in (b) shows a temporal evolution of the intrinsic $DOP$. $E_{det} = 799.4$ meV, $T = 4.2$ K, $P_{exc} = 30$ $\mu$W.

In order to study a temporal evolution of the $DOP$, additional two TRPL traces were measured in the following excitation-detection configurations: V-H, and H-V. The obtained $DOP$ is presented in Fig. 3(b) (black squares)—for the $DOP_V$, and Fig. 3(c) (red circles)—for the $DOP_H$. The asymmetry between $DOP_H$ and $DOP_V$ discussed for time-integrated spectra is naturally present here as well. Unfortunately, the amplitude of the $DOP_H$ is so small that it is registered with a rather large uncertainty in the present experiment. The $DOP_V$ decays from ~50% to ~27% within a time interval of 4.5 ns, which gives ~1.7 ns of the decay time constant. This time is not strictly related to the spin polarization lifetime since it results mainly from the difference in oscillator strengths for H and V directions. However, the observation of purely monoexponential decays (up to the ~6 ns time interval) allows us to estimate, basing on a rate equation model, that any spin relaxation process present in the system has to be slower than 15 ns.

In conclusion, we investigated the impact of various schemes of the optical spin pumping in the self-assembled InAs/In$_{0.53}$Ga$_{0.23}$Al$_{0.24}$As/InP(001) quantum dash structure emitting at 1.55 $\mu$m on spin memory of the ground state. The highly non-resonant spin pumping did not lead to preservation of the spin memory of the excitation, however, the registered polarization degree of ~25% pointed at an important intrinsic property of QDashes caused by the strong valence band mixing and anisotropic exchange interaction. In the case of spin injection into the wetting layer, the $DOP$ increased considerably by more than 15% with respect to the intrinsic $DOP$, which means the presence of spin memory effect. Furthermore, the best results were achieved in the case of exciton-1LO pumping scheme, for which the resultant $DOP$ exceeded ~50%.

This research was supported by The National Science Center Grant MAESTRO No. 2011/02/A/ST3/00152. The authors acknowledge the technical support from Hamamatsu Co. in near infrared streak camera apparatus. M.G. would like to thank Professor P. Machnikowski for valuable discussions. Ł.D. acknowledges the financial support from the Foundation for Polish Science within the START fellowship.

[1]J. Brault, M. Gendry, G. Grenet, G. Hollinger, Y. Desiéres, and T. Benyattou, Appl. Phys. Lett. **73**, 2932 (1998).
[2]H. Li, T. Daniel-Rice, and M.-A. Hasan, Appl. Phys. Lett. **80**, 1367 (2002).
[3]A. Sauerwald, T. Kümmell, G. Bacher, A. Somers, R. Schwertberger, J. P. Reithmaier, and A. Forchel, Appl. Phys. Lett. **86**, 253112 (2005).
[4]J. P. Reithmaier, A. Somers, S. Deubert, R. Schwertberger, W. Kaiser, A. Forchel, M. Calligaro, P. Resneau, O. Parillaud, S. Bansropun, M. Krakowski, R. Alizon, D. Hadass, A. Bilenca, H. Dery, V. Mikhelashvili, G. Eisenstein, M. Gioannini, I. Montrosset, T. W. Berg, M. van der Poel, J. Mørk, and B. Tromborg, J. Phys. D: Appl. Phys. **38**, 2088 (2005).
[5]M. Z. M. Khan, T. K. Ng, and B. S. Ooi, Prog. Quantum Electron. **38**, 237 (2014).
[6]F. Lelarge, B. Dagens, J. Renaudier, R. Brenot, A. Accard, F. van Dijk, D. Make, O. Le Gouezigou, J.-G. Provost, F. Poingt, J. Landreau, O. Drisse, E. Derouin, B. Rousseau, F. Pommereau, and G.-H. Duan, J. Sel. Top. Quantum Electron. **13**, 111 (2007).
[7]J. P. Reithmaier, G. Eisenstein, and A. Forchel, Proc. IEEE **95**, 1779 (2007).
[8]Ł. Dusanowski, M. Syperek, P. Mrowiński, W. Rudno-Rudziński, J. Misiewicz, A. Somers, S. Höfling, M. Kamp, J. P. Reithmaier, and G. Sęk, Appl. Phys. Lett. **105**, 021909 (2014).
[9]P. Mrowiński, A. Musiał, A. Maryński, M. Syperek, J. Misiewicz, A. Somers, J. P. Reithmaier, S. Höfling, and G. Sęk, Appl. Phys. Lett. **106**, 053114 (2015).
[10]W. Sheng and P. Hawrylak, Phys. Status Solidi C **3**, 3744 (2006).
[11]N. A. J. M. Kleemens, J. van Bree, M. Bozkurt, P. J. van Veldhoven, P. A. Nouwens, R. Nötzel, A. Yu. Silov, P. M. Koenraad, and M. E. Flatté, Phys. Rev. B **79**, 045311 (2009).
[12]D. Kim, W. Sheng, P. J. Poole, D. Dalacu, J. Lefebvre, J. Lapointe, M. E. Reimer, G. C. Aers, and R. L. Williams, Phys. Rev. B **79**, 045310 (2009).
[13]V. V. Belykh, A. Greilich, D. R. Yakovlev, M. Yacob, J. P. Reithmeier, M. Benyoucef, and M. Bayer, Phys. Rev. B **92**, 165307 (2015).
[14]*Single Quantum Dot: Fundamentals, Applications, and New Concepts*, edited by P. Michler (Springer-Verlag, Berlin, 2003).
[15]R. Seguin, A. Schliwa, S. Rodt, K. Pötschke, U. W. Pohl, and D. Bimberg, Phys. Rev. Lett. **95**, 257402 (2005).
[16]L. He, M. Gong, Ch.-F. Li, G.-C. Guo, and A. Zunger, Phys. Rev. Lett. **101**, 157405 (2008).
[17]Ch.-H. Lin, W.-T. You, H.-Y. Chou, S.-J. Cheng, S.-D. Lin, and W.-H. Chang, Phys. Rev. B **83**, 075317 (2011).
[18]A. Musiał, P. Kaczmarkiewicz, G. Sęk, P. Podemski, P. Machnikowski, and J. Misiewicz, Phys. Rev. B **85**, 035314 (2012).
[19]P. Kaczmarkiewicz, A. Musiał, G. Sęk, P. Podemski, P. Machnikowski, and J. Misiewicz, Acta. Phys. Pol., A **119**, 633 (2011).
[20]R. Winkler, *Spin-Orbit Coupling Effects in Two-Dimensional Electron and Hole Systems*, Springer Tracts in Modern Physics Vol. 191 (Springer, Berlin, 2003).
[21]*Optical Orientation*, Modern Problems in Condensed Matter Science Vol. 8, edited by F. Meier and B. Zakharchenya (North-Holland, 1984).
[22]J. Andrzejewski, G. Sęk, E. OReilly, A. Fiore, and J. Misiewicz, J. Appl. Phys. **107**, 073509 (2010).
[23]M. Syperek, Ł. Dusanowski, J. Andrzejewski, W. Rudno-Rudziński, G. Sęk, J. Misiewicz, and F. Lelarge, Appl. Phys. Lett. **103**, 083104 (2013).